\documentclass[twocolumn,showpacs,preprintnumbers,amsmath,amssymb]{revtex4}
\usepackage{graphicx}
\usepackage{dcolumn}
\usepackage{bm}

\begin{document}
\title{Technicolor corrections to $b\bar{b} \rightarrow W^{\pm}\pi^{\mp}_t$ at the CERN Large Hadron Collider}
\author{Jinshu Huang}
\email{jshuang@foxmail.com} \affiliation{College of Physics $\&$
Electronic Engineering, Nanyang Normal University, Nanyang 473061,
People's Republic of China; \\ College of Physics $\&$ Information
Engineering, Henan Normal University, Xinxiang 453007, People's
Republic of China}
\author{Qunna Pan}
\author{Taiping Song}
\affiliation{College of Physics $\&$ Electronic Engineering, Nanyang
Normal University, Nanyang 473061, People's Republic of China}
\author{Gongru Lu}
\email{lugongru@sina.com} \affiliation{College of Physics $\&$
Information Engineering, Henan Normal University, Xinxiang 453007,
People's Republic of China}

\date{\today}

\begin{abstract}
In this paper we calculate the technicolor correction to the
production of a charged top pion in association with a $W$ boson via
$b\bar{b}$ annihilation at the CERN Large Hadron Collider in the
context of the topcolor assisted technicolor model. We find that the
cross section of $pp \rightarrow b\bar{b} \rightarrow
W^{\pm}\pi_t^{\mp}$ at the tree level can reach a few hundred
femtobarns for reasonable ranges of the parameters, roughly
corresponding to the result of the process $pp \rightarrow b\bar{b}
\rightarrow W^{\pm}H^{\mp}$ in the minimal supersymmetric standard
model; the relative corrections arising from the one-loop diagrams
are about a few percent to two dozen percent, and they will increase
 the cross section at the tree level. As a comparison, we
also discuss the size of the hadron cross section via the other
subprocess $gg \rightarrow W^{\pm}\pi_t^{\mp}$.
\end{abstract}


\maketitle

\section{\label{sec:level1}Introduction}

Technicolor theory \cite{Weinberg1976,Dimopoulos1979} is one of the
important candidates to probe new physics beyond the standard model
(SM), especially topcolor assisted technicolor (TC2) model proposed
by C. T. Hill \cite{Hill1995} --- it combines technicolor with
topcolor, with the former mainly responsible for electroweak
symmetry breaking (EWSB) and the latter for generating a major part
of the top quark mass. If technicolor is actually responsible for
EWSB, there are strong phenomenological arguments that its energy
scale is at most a few hundred GeV and that the lightest technicolor
pions are within reach of the ATLAS and CMS experiments at the Large
Hadron Collider (LHC) \cite{Carena2008}. The TC2 model predicts
three top pions ($\pi^0_t, \pi^{\pm}_t$), one top Higgs ($h^0_t$)
and the new gauge bosons ($Z', B$) with large Yukawa couplings to
the third generation quarks, so these new particles can be regarded
as a typical feature of this model.  Lots of signals of this model
have already been studied in the work environment of linear
colliders and hadron-hadron colliders
\cite{Wang2002,Yue2003,Cao2003}, but most of the attention was
focused on the neutral top pion and new gauge bosons. Here we wish
to discuss the prospects of charged top pions.

The search for Higgs bosons and new physics particles and the study
of their properties are among the prime objectives of the LHC.
Recently, lots of studies about the neutral Higgs production at the
LHC have been finished \cite{Jakobs2009}. For the production of
charged Higgs boson in association with a $W$ boson in the minimal
supersymmetric standard model, Ref. \cite{Dicus1989} investigates
$b\bar{b} \rightarrow W^{\pm}H^{\mp}$ at the tree level and $gg
\rightarrow W^{\pm}H^{\mp}$ at one loop. The electroweak corrections
and QCD corrections to $b\bar{b} \rightarrow W^{\pm}H^{\mp}$ have
already been calculated in Ref. \cite{Yang2000}, which shows that a
favorable scenario for $W^{\pm} H^{\mp}$ associated production would
be characterized by the conditions that $m_H >m_t-m_b$ and that
$\tan\beta$ are either close to unity or of order $m_t/m_b$, then
the $H^{\pm}$ bosons could not spring from on-shell top quarks and
could be so copiously produced at hadron colliders. The authors of
Ref. \cite{Huang2004} have already studied the process of
$W^{\pm}\pi^{\mp}_t$ associated production via $b\bar{b}$
annihilation at the tree level, which shows that the total cross
section $\sigma(p\bar{p} \rightarrow b\bar{b} \rightarrow W^{\pm}
\pi^{\mp}_t)$ is rather large when $\pi^{\pm}_t$ is not very heavy.
In this paper we shall discuss the production of top pions
$\pi_t^{\pm}$ in association with SM gauge bosons $W^{\mp}$
including the contributions arising from top pions $\pi^0_t,
\pi^{\pm}_t$ and top Higgs $h^0_t$, and calculate the total cross
section to one-loop order, to search for new physics particles and
test the TC2 model.

This paper is organized as follows.  Section II is devoted to our
analytical results of the cross section of $pp \rightarrow b\bar{b}
\rightarrow W^{\pm}\pi_t^{\mp}$ in terms of the well-known standard
notation of one-loop Feynman integrals. The numerical results and
conclusions are included in Sec. III.

\section{\label{sec:level2} The calculations of $\sigma(p\bar{p} \rightarrow b\bar{b} \rightarrow
W^{\pm} \pi^{\mp}_t)$ at one-loop level}

\begin{figure*}

\vspace{-1.5cm}

\includegraphics{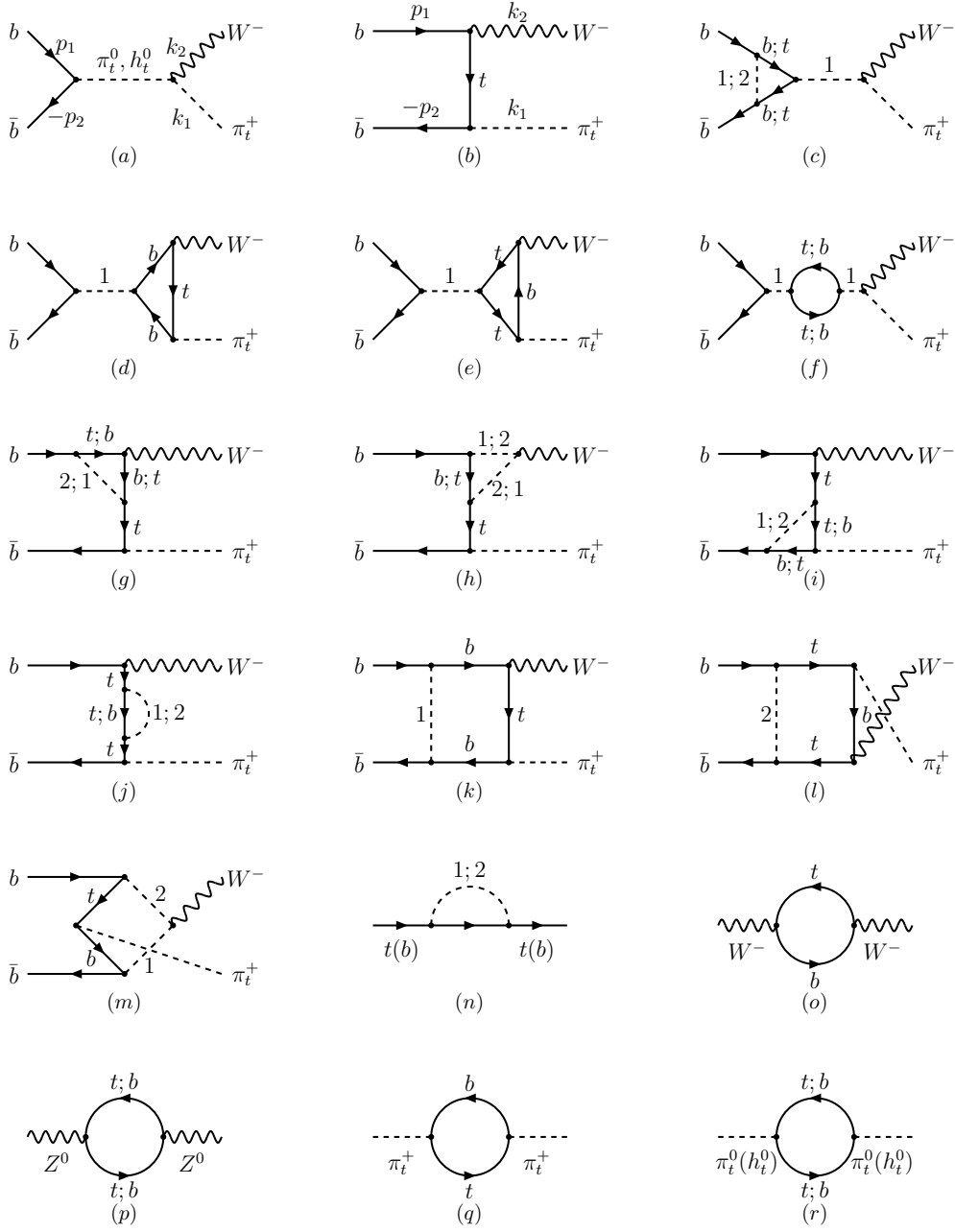}

\vspace{-11cm}

\caption{\label{fig:eps1} Feynman diagrams for the technicolor
corrections to the $b\bar{b} \rightarrow W^- \pi^+_t $ process:
(a)-(b) tree-level diagrams; (c)-(m) one-loop diagrams; (n)-(r) the
diagrams contributing to renormalization constants. Here the
internal dashed line 1 represents the neutral top pion $\pi^0_t$ and
top Higgs $h^0_t$, and the dashed line 2 denotes the charged top
pion $\pi^+_t$.}
\end{figure*}

The Feynman diagrams for the charged top pion production via $b(p_1)
\bar{b} (p_2) \rightarrow W^{\pm} (k_2) \pi^{\mp}_t (k_1)$, which
include the technicolor corrections to the process are shown in Fig.
\ref{fig:eps1}. The relevant Feynman rules are given in Refs.
\cite{Hill1995,Kaul1983}. In our calculation, we adopt the 't
Hooft-Feynman gauge and use the dimensional reduction for
regularization of the ultraviolet divergences in the virtual loop
corrections by the on-mass-shell renormalization scheme
\cite{Bohm1986}, in which the fine-structure constant $\alpha_{\rm
em}$ and physical masses are chosen to be the renormalized
parameters, and finite parts of the counterterms are fixed by the
renormalization conditions. The coupling constant $g$ is related to
the SM input parameters $e$, $m_W$ and $m_Z$.

The Mandelstam variables are defined as
\begin{eqnarray}
\hat{s}=(p_1+p_2)^2=(k_1+k_2)^2, \nonumber \\
\hat{t}=(p_1-k_2)^2=(p_2-k_1)^2, \nonumber \\
\hat{u}=(p_1-k_1)^2=(p_2-k_2)^2.
\end{eqnarray}

The self-energy coupling between $W^{+}$ and $\pi^{+}_t$ has no
contribution to their renormalization fields and masses since
$\pi^{+}_t$ is a pseudo- Goldstone boson and $W^+$ is a gauge boson.
We follow the approach of A. Mendez and A. Pomarol \cite{Mendez1992}
to define the relevant renormalization constants, the renormalized
amplitude for $b\bar{b} \rightarrow W^-\pi^+_t$ can be written as
\begin{equation}
M_{\rm rem}=M_0^{(\hat{s})}+M_0^{(\hat{t})}+\delta M,
\end{equation}
where $M^{(\hat{s})}_0$ and $M^{(\hat{t})}_0$ are the tree-level
amplitudes arising from Figs. 1 (a) and 1 (b), respectively, which
are given by
\begin{eqnarray}
M^{(\hat{s})}_0 &=& \frac{i g m_b
(1-\varepsilon)}{\sqrt{2}f_{\pi_t}}\frac{1}{\hat{s}-m^2_{\pi_t}}
\Big [\bar{v}(p_2)\gamma_5u(p_1) (p_1+p_2)_{\mu}
\nonumber \\
&& \times \varepsilon^{\mu}(k_2) \Big  ]  -\frac{i g
m_b(1-\varepsilon)}{\sqrt{2}f_{\pi_t}}\frac{1}{\hat{s}-m^2_{h_t}}
\Big  [\bar{v}(p_2) u(p_1)  \nonumber \\
&& \times (p_1+p_2)_{\mu}\varepsilon^{\mu}(k_2) \Big  ],
\end{eqnarray}
\begin{eqnarray}
M^{(\hat{t})}_0 &=& -\frac{i g (1-\varepsilon)}{\sqrt{2}f_{\pi_t}}\frac{1}{\hat{t}-m^2_{t}}
\Big  [m_b \bar{v}(p_2)(\not p_1-\not k_2) \gamma_{\mu} L u(p_1) \nonumber \\
&& \times \varepsilon^{\mu}(k_2)+m_t^2 \bar{v}(p_2)\gamma_{\mu} L
u(p_1) \varepsilon^{\mu}(k_2) \Big  ],
\end{eqnarray}
and  $\delta M$ denotes all of the one-loop corrections to the
tree-level process, which can be represented by
\begin{eqnarray}
\delta M&=& \Big  [\delta \hat{M}^{V_{1}(\hat{s})}+\delta
\hat{M}^{V_{2}(\hat{s})}+\delta \hat{M}^{S(\hat{s})} \Big  ](\pi^0_t) \nonumber \\
&& + \Big  [\delta \hat{M}^{V_{1}(\hat{s})}+\delta
\hat{M}^{V_{2}(\hat{s})}+\delta \hat{M}^{S(\hat{s})} \Big  ](h^0_t)
\nonumber \\
&& +\delta \hat{M}^{V_{1}(\hat{t})}+\delta \hat{M}^{V_{2}(\hat{t})}+\delta \hat{M}^{S(\hat{t})}+\delta M^B.
\end{eqnarray}

The amplitudes $\delta \hat{M}^{V}$ and $\delta \hat{M}^{S}$ arising
from the self-energy and vertex corrections can be written by
\begin{widetext}
\begin{eqnarray}
\delta \hat{M}^{V_1(\hat{s})} (\pi_t^0) &=& \frac{i g m_b
(1-\varepsilon)}{\sqrt{2}f_{\pi_t}}\frac{1}{\hat{s} -m^2_{\pi_t}}
\bar{v}(p_2) \bigg [\frac{\delta m_b}{m_b}+\frac{1}{2}\delta Z^b_L
+\frac{1}{2}\delta Z_R^b
+\frac{1}{2}\delta Z_{\pi^0_t} \bigg ]\gamma_5 (p_1+p_2)_{\mu} u(p_1) \varepsilon^{\mu}(k_2)\nonumber \\
&&  +\delta M^{V_1(\hat{s})} (\pi_t^0),
\end{eqnarray}
\begin{eqnarray}
\delta \hat{M}^{V_1(\hat{s})} (h_t^0) &=& -\frac{i g m_b
(1-\varepsilon)}{\sqrt{2}f_{\pi_t}}\frac{1}{\hat{s}
-m^2_{h_t}}\bar{v}(p_2) \bigg [\frac{\delta
m_b}{m_b}+\frac{1}{2}\delta Z^b_L +\frac{1}{2}\delta
Z_R^b+\frac{1}{2}\delta Z_{h^0_t} \bigg ] (p_1+p_2)_{\mu} u(p_1) \varepsilon^{\mu}(k_2) \nonumber \\
&&  +\delta M^{V_1(\hat{s})} (h_t^0),
\end{eqnarray}
\begin{eqnarray}
\delta \hat{M}^{V_2(\hat{s})} (\pi_t^0) &=& \frac{i g m_b
(1-\varepsilon)}{\sqrt{2}f_{\pi_t}}\frac{1}{\hat{s}
-m^2_{\pi_t}}\bar{v}(p_2) \bigg [\frac{\delta
g}{g}+\frac{1}{2}\delta Z_W +\frac{1}{2}\delta Z_{\pi^0_t}
+\frac{1}{2}\delta Z_{\pi^{+}_t} \bigg ] \gamma_5 (p_1+p_2)_{\mu} u(p_1) \varepsilon^{\mu}(k_2) \nonumber \\
&&  +\delta M^{V_2(\hat{s})} (\pi_t^0),
\end{eqnarray}
\begin{eqnarray}
\delta \hat{M}^{V_2(\hat{s})} (h_t^0) &=& -\frac{i g m_b (1-\varepsilon)}{\sqrt{2}f_{\pi_t}}\frac{1}{\hat{s}
-m^2_{h_t}}\bar{v}(p_2) \left [\frac{\delta g}{g}+\frac{1}{2}\delta Z_W +\frac{1}{2}\delta Z_{h^0_t}
+\frac{1}{2} \delta Z_{\pi^{+}_t} \right ] (p_1+p_2)_{\mu}u(p_1) \varepsilon^{\mu}(k_2) \nonumber \\
&& +\delta M^{V_2(\hat{s})} (h_t^0),
\end{eqnarray}
\begin{eqnarray}
\delta \hat{M}^{S(\hat{s})} (\pi_t^0) &=& \frac{i g m_b
(1-\varepsilon)}{\sqrt{2}f_{\pi_t}}\frac{1}{(\hat{s}
-m^2_{\pi_t})^2}\bar{v}(p_2) \bigg [\delta
m^2_{\pi_t}+(m^2_{\pi_t}-\hat{s})\delta Z_{\pi^0_t} \bigg ]\gamma_5
(p_1+p_2)_{\mu} u(p_1) \varepsilon^{\mu}(k_2)
\nonumber \\
&& +\delta M^{S(\hat{s})} (\pi_t^0),
\end{eqnarray}
\begin{eqnarray}
\delta \hat{M}^{S(\hat{s})} (h_t^0) &=& -\frac{i g m_b
(1-\varepsilon)}{\sqrt{2}f_{\pi_t}}\frac{1}{(\hat{s}
-m^2_{h_t})^2}\bar{v}(p_2) \bigg [\delta
m^2_{h_t}+(m^2_{h_t}-\hat{s})\delta Z_{h^0_t} \bigg
](p_1+p_2)_{\mu} u(p_1) \varepsilon^{\mu}(k_2)
\nonumber \\
&& +\delta M^{S(\hat{s})} (h_t^0),
\end{eqnarray}
\begin{eqnarray}
\delta \hat{M}^{V_1(\hat{t})} &=& -\frac{i g
(1-\varepsilon)}{\sqrt{2}f_{\pi_t}}\frac{1}{\hat{t} -m^2_t} \bigg [
m_b \bar{v}(p_2) \Big  ( \frac{\delta g}{g}+\frac{1}{2}\delta Z^t_L
+\frac{1}{2}\delta Z^b_L +\frac{1}{2}\delta Z_W \Big  )(\not
p_1-\not k_2)\gamma_{\mu} L u(p_1)\varepsilon^{\mu}(k_2)
 \nonumber \\
&& +m^2_t \bar{v}(p_2) \Big  ( \frac{\delta g}{g}+\frac{1}{2}\delta
Z^t_L +\frac{1}{2}\delta Z^b_L +\frac{1}{2}\delta Z_W \Big  )
\gamma_{\mu} L u(p_1)\varepsilon^{\mu}(k_2) \bigg ] +\delta
M^{V_1(\hat{t})},
\end{eqnarray}
\begin{eqnarray}
\delta \hat{M}^{V_2(\hat{t})} &=& -\frac{i g
(1-\varepsilon)}{\sqrt{2}f_{\pi_t}}\frac{1}{\hat{t} -m^2_t} \bigg [
m_b \bar{v}(p_2) \Big  ( \frac{\delta m_b}{m_b}+\frac{1}{2}\delta
Z^b_R +\frac{1}{2}\delta Z^t_L +\frac{1}{2}\delta Z_{\pi^+_t} \Big )
(\not p_1-\not k_2)\gamma_{\mu} L u(p_1)\varepsilon^{\mu}(k_2)
+m^2_t \bar{v}(p_2) \nonumber \\
&& \times \Big  ( \frac{\delta m_t}{m_t}+\frac{1}{2}\delta Z^b_L
+\frac{1}{2}\delta Z^t_R +\frac{1}{2}\delta Z_{\pi^+_t} \Big  )
\gamma_{\mu} L u(p_1)\varepsilon^{\mu}(k_2) \bigg ] +\delta
M^{V_2(\hat{t})},
\end{eqnarray}
\begin{eqnarray}
\delta \hat{M}^{S(\hat{t})} &=& \frac{i g
(1-\varepsilon)}{\sqrt{2}f_{\pi_t}}\frac{1}{(\hat{t} -m^2_t)^2}
\bigg [ m_b \bar{v}(p_2)\delta Z^t_L (\not p_1-\not k_2)(\not
p_1-\not k_2)(\not p_1 -\not k_2) \gamma_{\mu}L
u(p_1)\varepsilon^{\mu}(k_2)+
\bar{v}(p_2) \Big ( \frac{1}{2}m^2_t \delta Z^t_L \nonumber \\
&& + \frac{1}{2}m^2_t \delta Z^t_R -m_t \delta m_t \Big  )  (\not
p_1 -\not k_2)(\not p_1-\not k_2) \gamma_{\mu}L
u(p_1)\varepsilon^{\mu}(k_2) + \bar{v}(p_2)\Big  ( m^2_t (m_t
-m_b)\delta Z^t_R-m_b m^2_t \delta Z^t_L \nonumber \\
&& - 2m_b m_t \delta m_t \Big  ) (\not p_1 -\not k_2) \gamma_{\mu}L
u(p_1)\varepsilon^{\mu}(k_2)
 -m^3_t \bar{v}(p_2) \Big  (\frac{1}{2}m_t \delta Z^t_L +\frac{1}{2}m_t \delta Z^t_R+ \delta m_t \Big
 )  \gamma_{\mu}L u(p_1)\varepsilon^{\mu}(k_2)  \bigg ] \nonumber \\
&& +\delta M^{S(\hat{t})}.
\end{eqnarray}
\end{widetext}

The one-loop amplitudes $\delta M^{V_{1}(\hat{s})} (\pi^0_t)$,
$\delta M^{V_{2}(\hat{s})} (\pi^0_t)$, $\delta M^{S(\hat{s})}
(\pi^0_t)$, $\delta M^{V_{1}(\hat{s})} (h^0_t)$, $\delta
M^{V_{2}(\hat{s})} (h^0_t)$, $\delta M^{S(\hat{s})} (h^0_t)$,
$\delta M^{V_{1} (\hat{t})}$, $\delta M^{V_{2}(\hat{t})}$, $\delta
M^{S(\hat{t})}$, and $\delta M^B$ represent the irreducible
corrections arising, respectively, from the $b\bar{b}\pi^0_t$ vertex
diagram shown in Fig. 1(c),  the $W^- \pi^+_t \pi^0_t$ vertex
diagrams in Figs. 1(d)-1(e), the $\pi^0_t$ self-energy diagram in
Fig. 1(f), the $b\bar{b}h^0_t$ vertex diagram in Fig. 1(c), the $W^-
\pi^+_t h^0_t$ vertex diagrams in Figs. 1(d)-1(e), the $h^0_t$
self-energy diagram in Fig. 1(f), the $btW^-$ vertex diagrams in
Figs. 1(g)-1(h), the $t\bar{b}\pi^+_t$ vertex diagrams in Fig. 1(i),
the top quark self-energy diagram in Fig. 1(j), and the box diagrams
in Figs. 1(k)-1(m).

Calculating the self-energy diagrams in Figs. 1(n)-1(r),  we can get
the expressions of all the renormalization constants.

The detailed expressions of all above $\delta M^{V}$, $\delta M^S$,
and $\delta M^B$  and the renormalization constants are tedious, so
we do not present them here.

The corresponding amplitude squared is
\begin{eqnarray}
\overline{\sum} \big |M_{\rm ren} \big |^2 &=& \overline{\sum} \big |M^{(\hat{s})}_0+M^{(\hat{t})}_0 \big |^2 \nonumber \\
&& + 2{\rm Re} \overline{\sum} \big  [\delta M \big
(M^{(\hat{s})}_0+M^{(\hat{t})}_0 \big )^{\dag} \big  ].
\end{eqnarray}

The cross section for the process $b\bar{b} \rightarrow
W^{\pm}\pi_{t}^{\mp}$ is
\begin{equation}
\hat{\sigma}=\int_{\hat{t}_-}^{\hat{t}_+}
\frac{1}{16\pi\hat{s}^2}\overline{\sum} \big |M_{\rm ren} \big
|^2{\rm d}\hat{t},
\end{equation}
with
\begin{eqnarray}
\hat{t}_{\pm} &=& \frac{m_W^2+m_{\pi_t}^2-\hat{s}}{2} \nonumber \\
&& \pm\frac{1}{2} \sqrt{ \left [\hat{s}-(m_W+m_{\pi_t})^2 \right ]
\left [\hat{s}-(m_W-m_{\pi_t})^2 \right ]}.  \nonumber \\ && \
\end{eqnarray}
The total hadronic cross section for $pp \rightarrow b\bar{b}
\rightarrow W^{\pm}\pi_t^{\mp}$  can be obtained by folding the
subprocess cross section $\hat{\sigma}$ with the parton luminosity
\cite{Dicus1989}
\begin{equation}
\sigma(s)=\int_{(m_W+m_{\pi_t})/\sqrt{s}}^1 {\rm d}z \frac{{\rm
d}L}{{\rm d} z}\hat{\sigma}(b\bar{b}\rightarrow W^{\pm}\pi_t^{\mp}\
{\rm at}\ \hat{s}=z^2 s).
\end{equation}
Here  $\sqrt{s}$ and $\sqrt{\hat{s}}$ are the c.m. energies of the
$pp$ and $b\bar{b}$ states,  respectively, and ${\rm d}L/{\rm d}z$
is the parton luminosity, defined as \cite{Dicus1989,Yang2000}
\begin{equation}
\frac{{\rm d}L}{{\rm d}z}=2z\int_{z^2}^{1}\frac{{\rm d}x}{x}f_{b/p}(x,\mu) f_{\bar{b}/p}(z^2/x,\mu),
\end{equation}
where $f_{b/p}(x,\mu)$ and $f_{\bar{b}/p}(z^2/x,\mu)$ are the bottom
quark and bottom antiquark  parton distribution functions,
respectively.

\section{\label{sec:level3} Numerical results and conclusions}

We are now in a position to explore the phenomenological
implications of our results.  The SM input parameters for our
numerical analysis are $G_F=1.16639 \times 10^{-5}\ {\rm GeV}^{-2}$,
$m_W=80.398\ {\rm GeV}$, $m_Z=91.1876\ {\rm GeV}$, $m_t=171.2\ {\rm
GeV}$, and $m_b=4.2\ {\rm GeV}$ \cite{Amsler2008}. We use LoopTools
\cite{Hahn1999} and the CTEQ6M parton distribution function
\cite{Pumplin2006} with $\mu=\sqrt{s}/2$. The parameter
$\varepsilon$ and the masses of top pion $\pi^0_t, \pi_t^{\pm}$ and
top Higgs $h^0_t$ are all model-dependent \cite{Hill1995}, we select
them as free parameters, and take
\begin{eqnarray}
0.03\leq \varepsilon \leq 0.1,\ 200\ {\rm GeV}\leq m_{\pi_t} \leq
600\ {\rm GeV},
\end{eqnarray} and $m_{h_t}= 150,\ 250\ {\rm GeV}$ to estimate the total cross section
of $W^{\pm}\pi_t^{\mp}$ associated production at the LHC. We sum
over the final states $W^+\pi^-_t$ and $W^-\pi^+_t$ considering
their symmetry.

\begin{figure}
\includegraphics{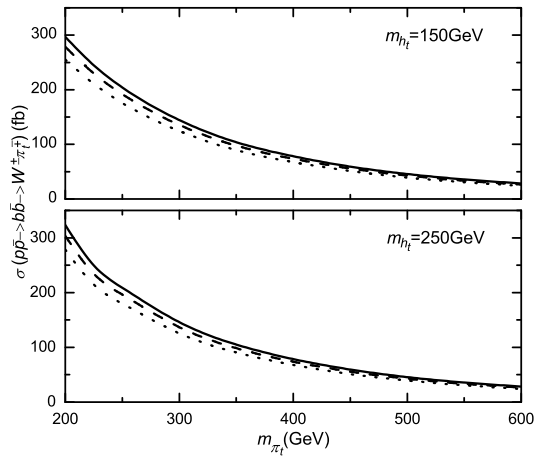}
\caption{\label{fig:eps2} The total cross section $\sigma(pp
\rightarrow b\bar{b}  \rightarrow W^{\pm}\pi_t^{\mp})$ at the tree
level versus  $m_{\pi_t}$ for $\varepsilon=0.03$ (solid line), 0.06
(dashed line), and 0.1 (dotted line).}
\end{figure}

\begin{figure}
\includegraphics{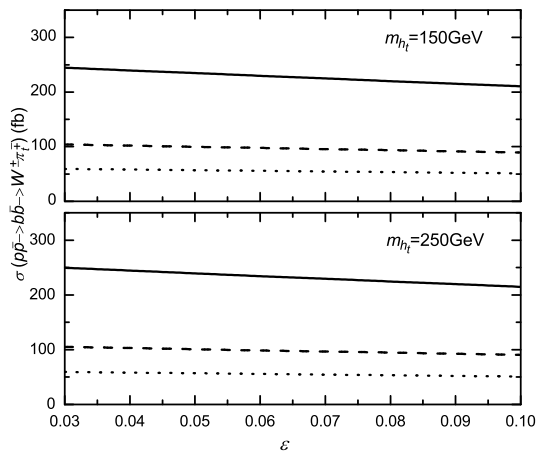}
\caption{\label{fig:eps3} The curve of $\sigma(pp \rightarrow
b\bar{b}  \rightarrow W^{\pm}\pi_t^{\mp})$  at the tree level vs.
$\varepsilon$ for $m_{\pi_t}=225\ {\rm GeV}$(solid line), 350 GeV
(dashed line), and 450 GeV (dotted line).}
\end{figure}

\begin{figure}
\includegraphics{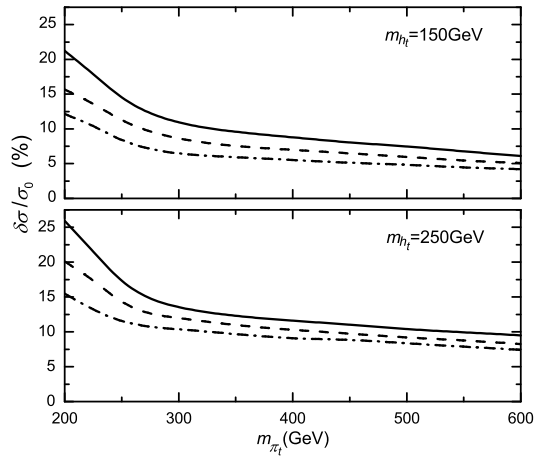}
\caption{\label{fig:eps4} The relative correction to the cross
section,  $\delta \sigma/\sigma_0$, as a function of $m_{\pi_t}$
with $m_{h_t}=150$, $250$ GeV and $\varepsilon = 0.03$ (solid line),
0.06 (dashed line), and 0.1 (dot-dashed line).}
\end{figure}

\begin{figure}
\includegraphics{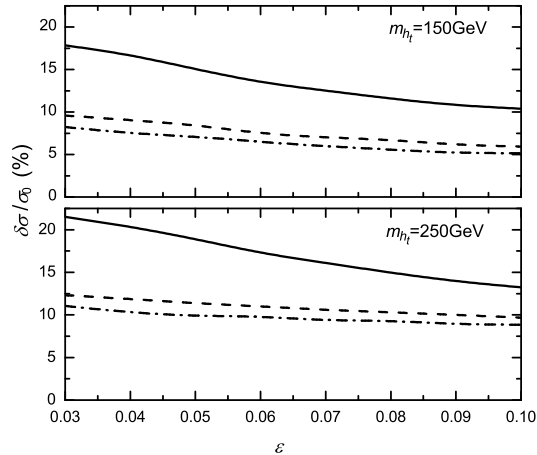}
\caption{\label{fig:eps5} The relative correction versus
$\varepsilon$  for $m_{\pi_t}=225\ {\rm GeV}$ (solid line), 350 GeV
(dashed line), and 450 GeV (dot-dashed line), respectively.}
\end{figure}

The final numerical results are summarized in Figs.
\ref{fig:eps2}-\ref{fig:eps5}.  In Fig. \ref{fig:eps2}, the total
cross section $\sigma(pp \rightarrow b\bar{b} \rightarrow
W^{\pm}\pi_t^{\mp})$  at the tree level as a function of $m_{\pi_t}$
for $m_{h_t}=150,\ 250\ {\rm GeV}$ at the LHC with $L=100\ {\rm
fb}^{-1}$ is given, in which the solid lines, dashed lines, and
dotted lines denote, respectively, the cases of $\varepsilon=0.03,
0.06$, and $0.1$. From this diagram, we can see that (i) the total
cross section decreases quickly as $m_{\pi_t}$ increase, changes the
values from $2.79 \times 10^2\ {\rm fb}$ to $26.6\ {\rm fb}$ with
the range of $m_{\pi_t}$, $200 \sim 600\ {\rm GeV}$ for
$\varepsilon=0.06$ and $m_{h_t}=150\ {\rm GeV}$, and from $3.04
\times 10^2\ {\rm fb}$ to $26.6\ {\rm fb}$ for $\varepsilon=0.06$
and $m_{h_t}=250\ {\rm GeV}$, respectively; (ii) the cross section
is sensitive to $\varepsilon$ and $m_{h_t}$ when $m_{\pi_t}$ is
small, but this sensitivity will disappear for a rather heavy top
pion; and (iii) when $m_{\pi_t}=225\ {\rm GeV}$, the cross section
of $W^{\pm}\pi_t^{\mp}$ associated production via $b\bar{b}$
annihilation is roughly $250\ {\rm fb}$, and is rather large.

Figure \ref{fig:eps3} gives the plots of the fully integrated cross
section  via $b\bar{b}$ annihilation at the tree level versus
$\varepsilon$ for $m_{\pi_t}=225,\ 350,$ and $450\ {\rm GeV}$. We
can observe that (i) the cross section is not sensitive to
$\varepsilon$, and only decreases by $13.6\% \sim 14.2\%$ in the
range of $0.03 \leq \varepsilon \leq 0.1$ for $m_{h_t}=150\ {\rm
GeV}$; and (ii) the case of $m_{h_t}=250\ {\rm GeV}$ is almost the
same as that of $m_{h_t}=150\ {\rm GeV}$.

In order to look at the contributions of the one-loop Feynman
diagrams,  we plot the relative correction $\delta \sigma /\sigma_0$
as a function of $m_{\pi_t}$ and $\varepsilon$ in Figs.
\ref{fig:eps4}-\ref{fig:eps5}, in which $\sigma_0$ denotes the
corresponding total cross section at the tree level. From these two
diagrams, we can find that (i) for the case of $m_{h_t}=150\ {\rm
GeV}$, the relative correction $\delta \sigma /\sigma_0$ is
positive, and decreases with the increases of $m_{\pi_t}$ and
$\varepsilon$, but this decrease is rapid with $m_{\pi_t}$ and is
slow with $\varepsilon$ at the changing from $0.03$ to $0.1$;  (ii)
for the case of $m_{h_t}=250\ {\rm GeV}$, the relative correction is
slightly larger than that of $m_{h_t}=150\ {\rm GeV}$; and (iii) the
value of the relative correction, in general, is about a few percent
to two dozen percent, this means that the contribution from the
one-loop corrections can afford a more distinct change than that of
the tree level.

As is known, at the LHC, the integrated luminosity is expected to
reach $L=100\ {\rm fb}^{-1}$ per year, this shows that a cross
section of $1\ {\rm fb}$ could translate into about 60 detectable
$W^{\pm}H^{\mp}$ events per year \cite{Dicus1989,Amsler2008}.
Looking at Figs. \ref{fig:eps2} and \ref{fig:eps3}, we thus conclude
that, if $m_{\pi_t}=225\ {\rm GeV}$, depending on $\varepsilon$, one
should be able to collect an annual total of between $1.26 \times
10^4$ and $1.5 \times 10^4$ events. So the $W^{\pm}\pi_t^{\mp}$
signal should be clearly visible at LHC unless $m_{\pi_t}$ is very
large. Moreover, the total cross section will enhance a few percent
to two dozen percent arising from the one-loop technicolor
corrections.

We know there are mainly two parton subprocesses that contribute to
the hadronic cross section $pp \rightarrow W^{\pm}\pi_t^{\mp}$: the
$b\bar{b}$ annihilation and the $gg$ fusion. In this paper, we only
focus on the discussion of the $b\bar{b}$ annihilation. For the $gg$
fusion, there is no tree-level contribution to the subprocess $gg
\rightarrow W^{\pm}\pi_t^{\mp}$ in the TC2 model, however, at the
one-loop level, the process $gg \rightarrow W^{\pm}\pi_t^{\mp}$ can
be induced by quark-loop diagrams including the triangle diagrams
and the box diagrams. Calculating the contributions of these
diagrams. we can find that, the cross section of  $pp \rightarrow gg
\rightarrow W^{\pm}\pi_t^{\mp}$ is between $11.3$ and $82.6\ {\rm
fb}$ for reasonable ranges of the parameters, and the changes of
$\sigma(pp \rightarrow gg \rightarrow W^{\pm}\pi_t^{\mp})$ with the
parameters $m_{\pi_t}$, $m_{h_t}$ and $\varepsilon$ are very similar
to those of $b\bar{b}$ annihilation at the tree level given in the
Figs. \ref{fig:eps2} and \ref{fig:eps3}. Whereas the production
cross section based on $gg \rightarrow W^{\pm}\pi_t^{\mp}$  can be
comparable to that via $b\bar{b} \rightarrow W^{\pm}\pi_t^{\mp}$ due
to the large number of gluons in the high energy proton beams at the
LHC. The total cross section of $W^{\pm}\pi_t^{\mp}$ associated
production at the LHC should be the sum over these two parton
subprocesses.

In conclusion, we have calculated the technicolor corrections to the
cross section for $W^{\pm}\pi_t^{\mp}$ associated production via
$b\bar{b}$ annihilation at the CERN LHC in the topcolor assisted
technicolor model. We find that, the total cross section of $pp
\rightarrow b\bar{b} \rightarrow W^{\pm}\pi_t^{\mp}$ at the tree
level, is roughly corresponding to that of the process $pp
\rightarrow b\bar{b} \rightarrow W^{\pm}H^{\mp}$ in the minimal
supersymmetric standard model, and can reach a few hundred
femtobarns with reasonable values of the parameters. Considering the
technicolor corrections arising from the one-loop diagrams, the
relative correction is about a few percent to two dozen percent. The
size of the cross section via the subprocess $gg \rightarrow
W^{\pm}\pi_t^{\mp})$ is calculated. Thus, it is so large that the
signal of charged top pion should be clearly visible at LHC.

\vspace{-3mm}

\section*{ACKNOWLEDGMENTS}

This project was supported in part by the National Natural Science
Foundation  of China under Grant Nos. 10975047 and 10979008; the
Natural Science Foundation of Henan Province under Nos. 092300410205
and 102300410210.

\end{document}